\documentclass{svproc}
\usepackage{url}
\usepackage{graphicx}
\usepackage{multicol}
\usepackage{footmisc}
\usepackage{amsmath}
\usepackage{amssymb}
\usepackage{color}

\usepackage{boondox-cal}
\newcommand{\zeroMatrix}{\pmb{0}}
\newcommand{\identityMatrix}{\pmb{I}}
\newcommand{\realNumbers}{\mathbb{R}}
\newcommand{\module}[1]{\left\|#1\right\|}
\newcommand{\abs}[1]{|#1|}

\newcommand{\minEigenvalue}[1]{\lambda_{\textrm{min}}(#1)}
\newcommand{\maxEigenvalue}[1]{\lambda_{\textrm{max}}(#1)}

\newcommand{\inertiaMatrix}{J}
\newcommand{\inertiaMatrixEstimate}{\hat{J}}
\newcommand{\configuration}{q}
\newcommand{\configurationDerivative}{\dot{q}}
\newcommand{\configurationDerivativeEstimate}{\hat{\dot{q}}}
\newcommand{\configurationSecondDerivative}{\ddot{q}}
\newcommand{\dynamicalPhenomena}{h}
\newcommand{\externalDisturbance}{\tau^*}
\newcommand{\modeledDynamics}{h_m}
\newcommand{\modeledDynamicsEstimate}{\hat{h}_m}
\newcommand{\controlSignal}{\tau}

\newcommand{\controlError}{e}

\newcommand{\controlErrorDerivative}{\dot{e}}
\newcommand{\controlErrorDerivativeEstimate}{\hat{\dot{e}}}
\newcommand{\controlErrorSecondDerivative}{\ddot{e}}
\newcommand{\desiredConfiguration}{q_d}

\newcommand{\desiredConfigurationSecondDerivative}{\ddot{q}_d}

\newcommand{\differentiableFunctionSet}[1]{\mathcal{C}^{#1}}
\newcommand{\totalDisturbance}{f}
\newcommand{\totalDisturbanceDerivative}{\dot{f}}
\newcommand{\totalDisturbanceSecondDerivative}{\ddot{f}}
\newcommand{\totalDisturbanceThirdDerivative}{\dddot{f}}
\newcommand{\totalDisturbanceEstimate}{\hat{f}}
\newcommand{\proportionalGain}{k_p}
\newcommand{\differentialGain}{k_d}

\newcommand{\extendedState}{\pmb{x}}
\newcommand{\extendedStateDerivative}{\dot{\pmb{x}}}
\newcommand{\extendedStateEstimate}{\hat{\pmb{x}}}
\newcommand{\extendedStateEstimateDerivative}{\dot{\hat{\pmb{x}}}}
\newcommand{\observerStateMatrix}[1]{\pmb{A}_#1}
\newcommand{\observerOutputMatrix}[1]{\pmb{c}_#1}
\newcommand{\observerInputMatrix}[1]{\pmb{b}_#1}
\newcommand{\extendedStateOutput}{y}
\newcommand{\observerOutput}{\hat{y}}
\newcommand{\measurementNoise}{w}
\newcommand{\observerGainVector}[1]{\pmb{l}_#1}

\newcommand{\observerBandwidth}{\omega_o}

\newcommand{\moreExtendedState}{\pmb{\chi}}
\newcommand{\moreExtendedStateDerivative}{\dot{\pmb{\chi}}}
\newcommand{\moreExtendedStateOutput}{\mathcal{y}}

\newcommand{\oscillatoryExtendedState}{\pmb{\chi}^*}
\newcommand{\oscillatoryExtendedStateDerivative}{\dot{\pmb{\chi}}^*}
\newcommand{\oscillatoryExtendedStateOutput}{\mathcal{y}^*}

\newcommand{\oscillatoryExtendedStateEstimate}{\hat{\pmb{\chi}}^*}
\newcommand{\oscillatoryExtendedStateEstimateDerivative}{\dot{\hat{\pmb{\chi}}}^*}
\newcommand{\oscillatoryExtendedObserverOutput}{\hat{\mathcal{y}}^*}
\newcommand{\oscillatoryExtendedObserverStateMatrix}{\pmb{A}_5^*}
\newcommand{\oscillatoryExtendedObserverDisturbanceInputMatrix}{\pmb{d}_5^*}

\newcommand{\extendedObserverOutput}{\hat{\mathcal{y}}}
\newcommand{\moreExtendedStateEstimateDerivative}{\dot{\hat{\pmb{\chi}}}}
\newcommand{\moreExtendedStateEstimate}{\hat{\pmb{\chi}}}

\newcommand{\oscillatoryTotalDisturbance}{f_o}
\newcommand{\oscillatoryTotalDisturbanceDerivative}{\dot{f}_o}
\newcommand{\oscillatoryTotalDisturbanceSecondDerivative}{\ddot{f}_o}
\newcommand{\disturbancePulsation}{\omega_r}
\newcommand{\disturbanceInputMatrix}[1]{\pmb{d}_#1}
\newcommand{\residualTotalDisturbance}{f_r}
\newcommand{\residualTotalDisturbanceDerivative}{\dot{f}_r}

\newcommand{\extendedStateObservationError}{\tilde{\pmb{x}}}
\newcommand{\extendedStateObservationErrorStateMatrix}{\pmb{H}_{\tilde{x}}}
\newcommand{\extendedStateObservationErrorDerivative}{\dot{\tilde{\pmb{x}}}}

\newcommand{\controlErrorDerivativeObservationError}{\tilde{\dot{e}}}
\newcommand{\totalDisturbanceObservationError}{\tilde{f}}
\newcommand{\errorDynamicsObservationErrorGain}{\pmb{k}}
\newcommand{\combinedControlError}{\pmb{\varepsilon}}
\newcommand{\combinedControlErrorDerivative}{\dot{\pmb{\varepsilon}}}
\newcommand{\combinedControlErrorStateMatrix}{\pmb{H}_\varepsilon}

\newcommand{\controlErrorLyapunovFunction}{V_\varepsilon}
\newcommand{\controlErrorLyapunovFunctionDerivative}{\dot{V}_\varepsilon}
\newcommand{\controlErrorMajorizationConstant}{\nu_\varepsilon}
\newcommand{\controlErrorEigenvalue}{\rho_\varepsilon}
\newcommand{\controlErrorConvergenceSpeed}{\gamma_\varepsilon}
\newcommand{\controlErrorLyapunovEquationSolution}{\pmb{P}_\varepsilon}

\newcommand{\observationErrorLyapunovFunction}{V_{\tilde{x}}}
\newcommand{\observationErrorLyapunovFunctionDerivative}{\dot{V}_{\tilde{x}}}
\newcommand{\observationErrorMajorizationConstant}{\nu_{\tilde{x}}}

\newcommand{\observationErrorConvergenceSpeed}{\gamma_{\tilde{x}}}
\newcommand{\observationErrorLyapunovEquationSolution}{\pmb{P}_{\tilde{x}}}

\newcommand{\errorCost}{J_e}
\newcommand{\controlCost}{J_u}
\newcommand{\disturbanceErrorCost}{J_f}

\newcommand{\generalizedStateVector}{\pmb{z}_n}
\newcommand{\generalizedStateVectorElement}[1]{z_#1}

\newcommand{\generalizedStateVectorElementDerivative}[1]{\dot{z}_#1}
\newcommand{\reducedGeneralizedStateVector}[1]{\pmb{z}_#1}

\newcommand{\reducedGeneralizedStateVectorEstimate}[1]{\hat{\pmb{z}}_#1}

\newcommand{\nonlinearStateFunctionElement}[1]{\phi_#1}

\newcommand{\nonlinearDisturbanceInputFunctionElement}[1]{g_#1}
\newcommand{\generalizedDisturbance}{\zeta}

\newcommand{\astolfiMarconiState}{\pmb{\xi}}
\newcommand{\astolfiMarconiSubState}[1]{\pmb{\xi}_{#1}}
\newcommand{\astolfiMarconiSubStateDerivative}[1]{\dot{\pmb{\xi}}_{#1}}

\newcommand{\astolfiMarconiGain}[1]{\pmb{\kappa}_{#1}}
\newcommand{\astolfiMarconiGainElement}[1]{\kappa_{#1}}
\newcommand{\astolfiMarconiError}[1]{\epsilon_{#1}}
\newcommand{\astolfiMarconiTransformationMatrix}{\pmb{L}}

\newcommand{\transferFunction}{G}
\newcommand{\laplaceOperator}{s}
\newcommand{\simulationTime}{T_{\textrm{sim}}}
\newcommand{\rank}{n}
\usepackage{xcolor,colortbl}
\usepackage{ragged2e}
\usepackage{wrapfig}

\definecolor{Gray}{gray}{0.85}
\definecolor{White}{rgb}{1,1,1}
\newcolumntype{a}{>{\columncolor{Gray}}c}
\newcolumntype{b}{>{\columncolor{white}}c}

\newcommand\blfootnote[1]{%
  \begingroup
  \renewcommand\thefootnote{}\footnote{#1}%
  \addtocounter{footnote}{-1}%
  \endgroup
}

\newcommand{\unit}{\ \mathrm}

\begin{document}
\mainmatter              % start of a contribution
%
% \title{A nonlinear ESO with limited gain power in the trajectory tracking ADR controller for a mechanical system: a case study}
\title{ESO architectures in the trajectory tracking ADR controller for a mechanical system: \newline a comparison}
\titlerunning{ESO architectures in the ADR controller for a mechanical system}  % abbreviated title (for running head)
\author{Krzysztof {\L}akomy \and Rados{\l}aw Patelski \and Dariusz Pazderski}
%
% \authorrunning{Ivar Ekeland et al.} % abbreviated author list (for running head)
% \tocauthor{Ivar Ekeland, Roger Temam, Jeffrey Dean, David Grove,
% Craig Chambers, Kim B. Bruce, and Elisa Bertino}
%
\institute{Poznań University of Technology, Piotrowo 3A, 60-965, Poznań, Poland,\\
\email{krzysztof.pi.lakomy@doctorate.put.poznan.pl, radoslaw.z.patelski@doctorate.put.poznan.pl}}

\maketitle              % typeset the title of the contribution

\vspace{-0.5cm}
\begin{abstract}
Proper operation of the Active Disturbance Rejection \linebreak (ADR) controller requires a precise determination of the so-called total disturbance affecting the considered dynamical system, usually estimated by the Extended State Observer (ESO). The observation quality of total disturbance has a significant impact on the control error values, making room for a potential improvement of control system performance using different structures of ESO. In this article, we provide a quantitative comparison between the Luenberger and Astolfi/Marconi (AM) observers designed for three different extended state representations and utilized in the trajectory tracking ADR controller designed for a mechanical system. Included results were obtained in the simple simulation case, followed by the experimental validation on the main axis of a telescope mount. \blfootnote{This work was partially supported by 33/32/SIGR/0003 and 09/93/PNCN/0429.}
\end{abstract}

%%%%%%%%%%%%%%%%%%%%%%%%%%%%%%%%%%%%%%%%%%%%%%%%%%%%%%%%%%%%%%%%%%
\section{Introduction}
Research interest in the Active Disturbance Rejection (ADR) control method, initially introduced in \cite{han2009}, has been gradually growing in recent years resulting in its frequent use in the domains of industrial- \cite{madonski2019,xue2017,patelski2019} and mobile robotics \cite{michalek2019}. Instead of relying on the precise model of the control object, the ADR method depends on a feedforward cancellation of the so-called total disturbance, usually estimated by the Extended State Observer (ESO) with a single additional state, see \cite{gao2006}. Since the control quality obtained with the ADR-based controller is highly dependent on the total disturbance estimation accuracy, we may improve control performance by choosing the structure of ESO that is more suitable to the expected type of disturbance. Instead of increasing the observer gains to estimate quickly varying disturbances, one can implement an ESO augmented by multiple states, see \cite{martinez2009}, or use a Resonant Extended State Observer (RESO) \cite{madonski2019} in the presence of total disturbance including an oscillatory component. An ESO is most commonly implemented according to a high-gain Luenberger observer design method \cite{khalil2014}, what usually results in a strong amplification of the sensor measurement noise. To decrease the impact of measurement noise Astolfi and Marconi introduced a limited gain power observer for the systems in the canonical observability form \cite{astolfi2015}, which was generalized to a wider class of objects in \cite{wang2017}.

In this article, we would like to compare the results obtained for three aforementioned observer structures, corresponding to a conventional ESO augmented by a single and multiple states, and a specific extended state representation utilized in RESO. Chosen structures will be implemented using two different observer architectures, i.e., a high-gain Luenberger observer and the observer proposed by Astolfi/Marconi (AM). Since most of the observer architectures are presented in separate articles describing the expected control performance with various methods of analysis, a quantitative comparison between different observer types with a clearly defined comparison criterion is necessary. In our considerations, we focus on a trajectory tracking control task performed by a mechanical system under the presence of the external perturbations and measurement noise. Conducted simulations were made on a second-order linear dynamics, while the experimental studies were made on the main axis of a telescope mount, which in general can be modeled with nonlinear differential equations.

\textbf{Notation}: Throughout this paper, we assume that $\hat{x}$ correspond to the estimate of signal $x$, by marking $x\in\differentiableFunctionSet{n}$ we assume that signal $x$ is at least $n$-times differentiable, $\pmb{A}\succ0$ means that matrix $\pmb{A}$ is positive definite matrix in the sense that $\pmb{x}^\top\pmb{A}\pmb{x}>0$ for any vector $\pmb{x}$, the norm of a matrix $\pmb{A}$  is defined as $\module{\pmb{A}}\triangleq\sup\{\module{\pmb{A}\pmb{x}}:\pmb{x}\in\realNumbers^n \ \textrm{and} \ \module{\pmb{x}}=1 \}$, $\minEigenvalue{\pmb{A}}$ and $\maxEigenvalue{\pmb{A}}$ are respectively the minimal and maximal eigenvalues of matrix $\pmb{A}$, while $\zeroMatrix$ and $\identityMatrix$ represent zero and identity matrices of the appropriate order.

%%%%%%%%%%%%%%%%%%%%%%%%%%%%%%%%%%%%%%%%%%%%%%%%%%%%%%%%%%%%%%%%%%
\section{Preliminaries}

Let us consider a mechanical system having a single degree of freedom with the dynamics expressed as
\begin{align}
    \inertiaMatrix(\configuration)\configurationSecondDerivative + \dynamicalPhenomena(\configuration,\configurationDerivative) + \modeledDynamics(\configuration,\configurationDerivative) + \externalDisturbance = \controlSignal,
    \label{eq:systemDynamics}
\end{align}
where $\configuration\in\realNumbers$ corresponds to a system configuration, $\inertiaMatrix(\configuration)$ is a system inertia, $\modeledDynamics(\configuration,\configurationDerivative)$ corresponds to a known (or modeled) part of the system dynamics, $\dynamicalPhenomena(\configuration,\configurationDerivative)$ aggregates the unmodeled dynamical phenomena, $\externalDisturbance\in\differentiableFunctionSet{3}$ is a bounded external disturbance with bounded derivatives, and $\controlSignal$ is a control signal.
To solve the trajectory tracking motion task, we first have to define the desired trajectory $\desiredConfiguration\in\differentiableFunctionSet{5}$ included in the control error definition $\controlError \triangleq \desiredConfiguration - \configuration$, which dynamics can be derived upon \eqref{eq:systemDynamics} and written down in a form
\begin{align}
    \controlErrorSecondDerivative = \desiredConfigurationSecondDerivative - \configurationSecondDerivative &= \underbrace{\desiredConfigurationSecondDerivative - \frac{1}{\inertiaMatrix}(\controlSignal-\modeledDynamics-\dynamicalPhenomena-\externalDisturbance) + \frac{1}{\inertiaMatrixEstimate}(\controlSignal-\modeledDynamicsEstimate)}_{\totalDisturbance (\desiredConfigurationSecondDerivative, \configuration, \configurationDerivative, \controlSignal, \externalDisturbance)} - \frac{1}{\inertiaMatrixEstimate}(\controlSignal-\modeledDynamicsEstimate)
    \label{eq:errorDynamics}
\end{align}
for $\inertiaMatrixEstimate:=\inertiaMatrixEstimate(\configuration)$ and $\modeledDynamicsEstimate:=\modeledDynamicsEstimate(\configurationDerivativeEstimate)$  being respectively the estimates of the system inertia and modeled dynamics, while $\totalDisturbance(\cdot)$ represents the lumped total-disturbance. According to equation \eqref{eq:errorDynamics}, we propose the trajectory tracking controller
\begin{align}
    \controlSignal \triangleq \modeledDynamicsEstimate + \inertiaMatrixEstimate(\totalDisturbanceEstimate + \proportionalGain\controlError + \differentialGain \controlErrorDerivativeEstimate),
    \label{eq:controlLaw}
\end{align}
where $\proportionalGain,\differentialGain>0$ are the gains of a PD controller.%, and $\totalDisturbanceEstimate$ is the estimate of a total disturbance.
After a substitution of control law \eqref{eq:controlLaw} into equation \eqref{eq:errorDynamics}, we obtain a closed-loop dynamics
\begin{align}
    \controlErrorSecondDerivative = -\proportionalGain\controlError - \differentialGain\controlErrorDerivative + \totalDisturbanceObservationError + \differentialGain\controlErrorDerivativeObservationError
    \label{eq:closedLoopDynamics}
\end{align}
that may be treated as a linear system perturbed by signals $\totalDisturbanceObservationError\triangleq\totalDisturbance - \totalDisturbanceEstimate$ and $\controlErrorDerivativeObservationError\triangleq\controlErrorDerivative-\controlErrorDerivativeEstimate$. According to \eqref{eq:closedLoopDynamics}, we may see that the control performance depends on the precise estimations of $\totalDisturbance$ and $\controlErrorDerivative$ that should be provided by the observer.

To design ESO, we firstly need to define an extended state vector $\extendedState \triangleq [\controlError \ \controlErrorDerivative \ \totalDisturbance]^\top$, expressed here in the error domain (see \cite{michalek2016}), which dynamics can be derived upon \eqref{eq:errorDynamics} and described with the state-space equations
\begin{align}
    \begin{cases}
        \extendedStateDerivative = \observerStateMatrix{3}\extendedState - \frac{1}{\inertiaMatrixEstimate}\observerInputMatrix{3}(\controlSignal-\modeledDynamicsEstimate) + \disturbanceInputMatrix{3}\totalDisturbanceDerivative\\
        \extendedStateOutput = \observerOutputMatrix{3}\extendedState + \measurementNoise
    \end{cases},
    \label{eq:extendedStateDynamics}
\end{align}
where $\measurementNoise$ is a bounded measurement noise, $\extendedStateOutput$ is a system output, while  $\observerStateMatrix{n}\triangleq\begin{bmatrix}\zeroMatrix^{n-1\times 1} & \identityMatrix^{n-1} \\ 0 & \zeroMatrix^{1\times n-1}\end{bmatrix}, \ \observerInputMatrix{n}\triangleq \begin{bmatrix} 0 \\ 1 \\ \zeroMatrix^{n-2\times 1} \end{bmatrix}, \ \observerOutputMatrix{n} \triangleq \begin{bmatrix} 1 & \zeroMatrix^{1\times n-1}\end{bmatrix}, \textrm{and} \ \disturbanceInputMatrix{n} \triangleq \begin{bmatrix} \zeroMatrix^{n-1}\times 1 \\ 1\end{bmatrix}.$ \linebreak
%
% \begin{align}
%     \observerStateMatrix{n}\triangleq \begin{bmatrix}\zeroMatrix^{n-1\times 1} & \identityMatrix^{n-1} \\ 0 & \zeroMatrix^{1\times n-1}\end{bmatrix}, \ \observerInputMatrix{n}\triangleq \begin{bmatrix} 0 \\ 1 \\ \zeroMatrix^{n-2\times 1} \end{bmatrix}, \ \observerOutputMatrix{n} \triangleq \begin{bmatrix} 1 & \zeroMatrix^{1\times n-1}\end{bmatrix}, \ \disturbanceInputMatrix{n} = \begin{bmatrix} \zeroMatrix^{n-1}\times 1 \\ 1\end{bmatrix}. \nonumber
% \end{align}
%
A standard Luenberger ESO estimating the values of $\extendedState$ and designed according to the equations \eqref{eq:extendedStateDynamics} is expressed as
\begin{align}
    \begin{cases}
        \extendedStateEstimateDerivative = \observerStateMatrix{3}\extendedStateEstimate - \frac{1}{\inertiaMatrixEstimate}\observerInputMatrix{3}(\controlSignal-\modeledDynamicsEstimate)+ \observerGainVector{3}(\observerOutput-\extendedStateOutput)\\
        \observerOutput = \observerOutputMatrix{3}\extendedStateEstimate
    \end{cases},
    \label{eq:standardESO}
\end{align}
where $\observerGainVector{3}\triangleq[3\observerBandwidth \ 3\observerBandwidth^2 \ \observerBandwidth^3]^\top$ is the observer gain vector dependent on a single parameter $\observerBandwidth$. The quality of estimation can be determined by analyzing the dynamics of the observation error $\extendedStateObservationError \triangleq \extendedState - \extendedStateEstimate$ derived upon \eqref{eq:extendedStateDynamics} and \eqref{eq:standardESO}, i.e.,
\begin{align}
    \extendedStateObservationErrorDerivative = \underbrace{(\observerStateMatrix{3}-\observerGainVector{3}\observerOutputMatrix{3})}_{\extendedStateObservationErrorStateMatrix}\extendedStateObservationError+\disturbanceInputMatrix{3}\totalDisturbanceDerivative+\observerGainVector{3}\measurementNoise .
    \label{eq:observationErrorDynamics}
\end{align}
Let us introduce a positive-definite function $\observationErrorLyapunovFunction \triangleq \frac{1}{2}\extendedStateObservationError^\top\observationErrorLyapunovEquationSolution\extendedStateObservationError$ bounded by \linebreak
$\frac{1}{2}\minEigenvalue{\observationErrorLyapunovEquationSolution}\module{\extendedStateObservationError}^2\leq\observationErrorLyapunovFunction\leq\frac{1}{2}\maxEigenvalue{\observationErrorLyapunovEquationSolution}\module{\extendedStateObservationError}^2$, where a symmetric matrix $\observationErrorLyapunovEquationSolution\succ0$ is a solution of Lyapunov equation $\extendedStateObservationErrorStateMatrix\observationErrorLyapunovEquationSolution+\observationErrorLyapunovEquationSolution\extendedStateObservationErrorStateMatrix^\top+\observerBandwidth\identityMatrix = \zeroMatrix$. The time derivative $\observationErrorLyapunovFunctionDerivative$ is bounded by
\begin{align}
    \observationErrorLyapunovFunctionDerivative &= \frac{1}{2}\extendedStateObservationError^\top(\extendedStateObservationErrorStateMatrix^\top\observationErrorLyapunovEquationSolution+\observationErrorLyapunovEquationSolution\extendedStateObservationErrorStateMatrix)\extendedStateObservationError + \extendedStateObservationError^\top\observationErrorLyapunovEquationSolution\disturbanceInputMatrix{3}\totalDisturbanceDerivative + \extendedStateObservationError^\top\observationErrorLyapunovEquationSolution\observerGainVector{3}\measurementNoise \nonumber \\
    &\leq -\frac{1}{2}(1-\observationErrorMajorizationConstant)\observerBandwidth\module{\extendedStateObservationError}^2 + \module{\extendedStateObservationError}\left(\module{\observationErrorLyapunovEquationSolution}\abs{\totalDisturbanceDerivative}+\observerBandwidth^3\module{\observationErrorLyapunovEquationSolution}\abs{\measurementNoise}- \frac{1}{2}\observationErrorMajorizationConstant\observerBandwidth\module{\extendedStateObservationError}\right)
\end{align}
%
% \begin{align}
%     \observationErrorLyapunovFunctionDerivative &= \frac{1}{2}\extendedStateObservationError^\top(\extendedStateObservationErrorStateMatrix^\top\observationErrorLyapunovEquationSolution+\observationErrorLyapunovEquationSolution\extendedStateObservationErrorStateMatrix)\extendedStateObservationError + \frac{1}{2}\totalDisturbanceDerivative\disturbanceInputMatrix{3}^\top\observationErrorLyapunovEquationSolution\extendedStateObservationError + \frac{1}{2}\extendedStateObservationError^\top\observationErrorLyapunovEquationSolution\disturbanceInputMatrix{3}\totalDisturbanceDerivative \nonumber \\
%     &+ \frac{1}{2}\measurementNoise\observerGainVector{3}^\top\observationErrorLyapunovEquationSolution\extendedStateObservationError + \frac{1}{2}\extendedStateObservationError^\top\observationErrorLyapunovEquationSolution\observerGainVector{3}\measurementNoise \nonumber \\
%     &\leq -\frac{1}{2}(1-\observationErrorMajorizationConstant)\observerBandwidth\module{\extendedStateObservationError}^2 + \module{\extendedStateObservationError}\left(\module{\observationErrorLyapunovEquationSolution}\abs{\totalDisturbanceDerivative}+\observerBandwidth^3\module{\observationErrorLyapunovEquationSolution}\abs{\measurementNoise}- \observationErrorMajorizationConstant\observerBandwidth\module{\extendedStateObservationError}\right)
% \end{align}
%
and fulfills the relation
\begin{align}
    \observationErrorLyapunovFunctionDerivative \leq  -\frac{1}{2}(1-\observationErrorMajorizationConstant)\observerBandwidth\module{\extendedStateObservationError}^2 \ \textrm{when} \ \module{\extendedStateObservationError} \geq \frac{2\module{\observationErrorLyapunovEquationSolution}}{\observationErrorMajorizationConstant\observerBandwidth}\abs{\totalDisturbanceDerivative}+\frac{2\observerBandwidth^2\module{\observationErrorLyapunovEquationSolution}}{\observationErrorMajorizationConstant}\abs{\measurementNoise}
    \label{eq:observationErrorLyapunovFunctionDerivativeResult}
\end{align}
for some majorization constant $\observationErrorMajorizationConstant\in(0,1)$. According to Th 5.1 from \cite{khalil2002}, multi-input ISS procedure utilized in \cite{lakomy2020} and \cite{peng2018}, and to the relation \eqref{eq:observationErrorLyapunovFunctionDerivativeResult}, a time response of the dynamics \eqref{eq:observationErrorDynamics} is bounded by
\begin{align}
    \forall_{t\geq 0} \module{\extendedStateObservationError(t)} &\leq c_1\module{\extendedStateObservationError(0)}e^{-\observationErrorConvergenceSpeed t}+\frac{2\module{\observationErrorLyapunovEquationSolution}}{\observationErrorMajorizationConstant\observerBandwidth}\sup_{t\geq0}\abs{\totalDisturbanceDerivative(t)}+\frac{2\observerBandwidth^2\module{\observationErrorLyapunovEquationSolution}}{\observationErrorMajorizationConstant}\sup_{t\geq0}\abs{\measurementNoise(t)},
    \label{eq:observationErrorISSResult}
\end{align}
where $c_1 = \sqrt{\maxEigenvalue{\observationErrorLyapunovEquationSolution}/\minEigenvalue{\observationErrorLyapunovEquationSolution}}$ and
$\observationErrorConvergenceSpeed = \observerBandwidth(1-\observationErrorMajorizationConstant)/(2\maxEigenvalue{\observationErrorLyapunovEquationSolution})$. In the result \eqref{eq:observationErrorISSResult} we can see two components involving the perturbation signals, where the first one depends on the derivative of total disturbance and can be reduced to arbitrarily small value as $\observerBandwidth\rightarrow\infty$, while the second one depending on the measurement noise $\measurementNoise(t)$ is amplified by the factor $\observerBandwidth^2$ and thus raise to infinity as $\observerBandwidth\rightarrow\infty$. In other words, there should exist the value $\observerBandwidth$ that minimizes the overall impact of perturbations on the observation error \eqref{eq:observationErrorISSResult}, as it was presented in the work \cite{khalil2014}.

Let us now consider a closed-loop dynamics of combined control error $\combinedControlError\triangleq[\controlError \ \controlErrorDerivative]^\top$ expressed as
\begin{align}
    \combinedControlErrorDerivative = \combinedControlErrorStateMatrix\combinedControlError+\observerInputMatrix{2}\errorDynamicsObservationErrorGain\extendedStateObservationError,
    \label{eq:combinedControlErrorDynamics}
\end{align}
%
% \begin{align}
%     \textrm{where} \quad \combinedControlErrorStateMatrix = \begin{bmatrix} 0 & 1 \\ -\proportionalGain & -\differentialGain \end{bmatrix} \quad \textrm{and}  \quad \errorDynamicsObservationErrorGain=[0 \ \differentialGain \ 1]. \nonumber
% \end{align}
%
where $\combinedControlErrorStateMatrix = \begin{bmatrix} 0 & 1 \\ -\proportionalGain & -\differentialGain \end{bmatrix}$  and  $\errorDynamicsObservationErrorGain=[0 \ \differentialGain \ 1]$.
Let us propose a positive-definite function $\controlErrorLyapunovFunction \triangleq \frac{1}{2}\combinedControlError^\top\controlErrorLyapunovEquationSolution\combinedControlError$ bounded by $\frac{1}{2}\minEigenvalue{\controlErrorLyapunovEquationSolution}\module{\combinedControlError}^2\leq\controlErrorLyapunovFunction\leq\frac{1}{2}\maxEigenvalue{\controlErrorLyapunovEquationSolution}\module{\combinedControlError}^2$, where a symmetric matrix $\controlErrorLyapunovEquationSolution\succ0$ is a solution of Lyapunov equation $\combinedControlErrorStateMatrix\controlErrorLyapunovEquationSolution+\controlErrorLyapunovEquationSolution\combinedControlErrorStateMatrix^\top+\controlErrorEigenvalue\identityMatrix = \zeroMatrix$ for some $\controlErrorEigenvalue>0$. The time derivative $\controlErrorLyapunovFunctionDerivative$ is bounded by
\begin{align}
    \controlErrorLyapunovFunctionDerivative &= \frac{1}{2}\combinedControlError^\top(\combinedControlErrorStateMatrix^\top\controlErrorLyapunovEquationSolution+\controlErrorLyapunovEquationSolution\combinedControlErrorStateMatrix)\combinedControlError + \combinedControlError^\top\controlErrorLyapunovEquationSolution\observerInputMatrix{2}\errorDynamicsObservationErrorGain\extendedStateObservationError \nonumber \\
    &\leq -\frac{1}{2}(1-\controlErrorMajorizationConstant)\controlErrorEigenvalue\module{\combinedControlError}^2 + \module{\combinedControlError}\left(\module{\controlErrorLyapunovEquationSolution}\module{\errorDynamicsObservationErrorGain}\module{\extendedStateObservationError} - \frac{1}{2}\controlErrorMajorizationConstant\controlErrorEigenvalue\module{\combinedControlError}\right)
\end{align}
and fulfills the relation
\begin{align}
    \controlErrorLyapunovFunctionDerivative \leq -\frac{1}{2}(1-\controlErrorMajorizationConstant)\controlErrorEigenvalue\module{\combinedControlError}^2  \ \textrm{when} \ \module{\combinedControlError}\geq\frac{2\module{\controlErrorLyapunovEquationSolution}\module{\errorDynamicsObservationErrorGain}}{\controlErrorMajorizationConstant\controlErrorEigenvalue}\module{\extendedStateObservationError}
    \label{eq:controlErrorLyapunovFunctionDerivativeResult}
\end{align}
for some majorization constant $\controlErrorMajorizationConstant\in(0,1)$. A time response of the dynamics \eqref{eq:combinedControlErrorDynamics}, according to the relation \eqref{eq:controlErrorLyapunovFunctionDerivativeResult}, is bounded by
\begin{align}
    \forall_{t\geq 0} \module{\combinedControlError(t)} \leq \sqrt{\frac{\maxEigenvalue{\controlErrorLyapunovEquationSolution}}{\minEigenvalue{\controlErrorLyapunovEquationSolution}}}\module{\combinedControlError(0)}e^{-\controlErrorConvergenceSpeed t}+\frac{2\module{\controlErrorLyapunovEquationSolution}\module{\errorDynamicsObservationErrorGain}}{\controlErrorMajorizationConstant\controlErrorEigenvalue}\sup_{t\geq0}\module{\extendedStateObservationError(t)},
    \label{eq:combinedControlErrorISSResult}
\end{align}
where $\controlErrorConvergenceSpeed = \controlErrorEigenvalue(1-\controlErrorMajorizationConstant)/(2\maxEigenvalue{\controlErrorLyapunovEquationSolution})$. The final value of control errors depends on the magnitude of observation error $\extendedStateObservationError(t)$, that according to \eqref{eq:observationErrorISSResult} is determined by the values of total disturbance derivative $\totalDisturbanceDerivative(t)$, the magnitude of a measurement noise $\measurementNoise(t)$, and the observation gain parameter $\observerBandwidth$. In the latter part of this article, we will introduce the observer architectures alternative to \eqref{eq:standardESO} that could possibly improve the estimation quality and noise attenuation.

% \begin{align}
%     \controlErrorSecondDerivative &= -\proportionalGain\controlErrorEstimate - \differentialGain\controlErrorDerivativeEstimate + \totalDisturbance - \totalDisturbanceEstimate \nonumber \\
%     &= -\proportionalGain\controlError - \differentialGain\controlErrorDerivative + \errorDynamicsObservationErrorGain\extendedStateObservationError
% \end{align}
% %
% $\errorDynamicsObservationErrorGain\triangleq[\proportionalGain \ \differentialGain \ 1]$

%%%%%%%%%%%%%%%%%%%%%%%%%%%%%%%%%%%%%%%%%%%%%%%%%%%%%%%%%%%%%%%%%%
\section{Alternative observer architectures}

To introduce the first of alternative observer structures, we need to assume that the total disturbance has a sinusoidal component, and can be rewritten as a sum
\begin{align}
    \totalDisturbance = \oscillatoryTotalDisturbance + \residualTotalDisturbance,
\end{align}
where $\oscillatoryTotalDisturbance$ refers to the oscillatory part of the disturbance, while $\residualTotalDisturbance$ contains the residual disturbances not included in $\oscillatoryTotalDisturbance$. We assume that the angular frequency of disturbance oscillations can be determined and is marked as $\disturbancePulsation$, thus we may model the resonant component of the total disturbance as a harmonic oscillator
\begin{align}
    \oscillatoryTotalDisturbanceSecondDerivative + \disturbancePulsation^2\oscillatoryTotalDisturbance = 0
    \label{eq:harmonicOscillator}
\end{align}
and consider it as a part of the new extended state dynamics
\begin{align}
    \begin{cases}
        \oscillatoryExtendedStateDerivative = \oscillatoryExtendedObserverStateMatrix\oscillatoryExtendedState - \frac{1}{\inertiaMatrixEstimate}\observerInputMatrix{5}(\controlSignal-\modeledDynamicsEstimate)+\oscillatoryExtendedObserverDisturbanceInputMatrix\residualTotalDisturbanceDerivative \\
        \oscillatoryExtendedStateOutput = \observerOutputMatrix{5}\oscillatoryExtendedState + \measurementNoise
    \end{cases},
    \label{eq:oscilatoryExtendedStateDynamics}
\end{align}
where
$\oscillatoryExtendedStateDerivative \triangleq [\controlError \ \controlErrorDerivative \ \totalDisturbance \ \oscillatoryTotalDisturbanceDerivative \ \oscillatoryTotalDisturbanceSecondDerivative]^\top$ is an extended state, $\oscillatoryExtendedStateOutput$ is an output of the system, while $  \oscillatoryExtendedObserverStateMatrix = \begin{bmatrix} \zeroMatrix^{4\times1} & & & \identityMatrix^4 \\
    0 & [0 & 0 & -\disturbancePulsation^2 & 0] \end{bmatrix}, \ \textrm{and} \ \oscillatoryExtendedObserverDisturbanceInputMatrix = [0 \ 0 \ 1 \ 0 \ 0]^\top.$
%
% \begin{align}
%     \oscillatoryExtendedObserverStateMatrix = \begin{bmatrix} \zeroMatrix^{4\times1} & & & \identityMatrix^4 \\
%     0 & [0 & 0 & -\disturbancePulsation^2 & 0] \end{bmatrix}, \ \oscillatoryExtendedObserverDisturbanceInputMatrix = [0 \ 0 \ 1 \ 0 \ 0]^\top.
%     % \begin{bmatrix} 0 & 1 & 0 & 0 & 0  \\
%     % 0 & 0 & 1 & 0 & 0 \\
%     % 0 & 0 & 0 & 1 & 0 \\
%     % 0 & 0 & 0 & 0 & 1 \\
%     % 0 & 0 & 0 & -\disturbancePulsation^2 & 0 \end{bmatrix}
% \end{align}
%
Based on the dynamics \eqref{eq:oscilatoryExtendedStateDynamics}, we can define the equations of RESO in a form
\begin{align}
    \begin{cases}
        \oscillatoryExtendedStateEstimateDerivative = \oscillatoryExtendedObserverStateMatrix\oscillatoryExtendedStateEstimate - \frac{1}{\inertiaMatrixEstimate}\observerInputMatrix{5}(\controlSignal-\modeledDynamicsEstimate)+ \observerGainVector{5}(\oscillatoryExtendedObserverOutput-\oscillatoryExtendedStateOutput)\\
        \oscillatoryExtendedObserverOutput = \observerOutputMatrix{3}\oscillatoryExtendedStateEstimate
    \end{cases},
    \label{eq:oscillatoryESO}
\end{align}
where $\observerGainVector{5}\triangleq[5\observerBandwidth \ 10\observerBandwidth^2 \ 10\observerBandwidth^3 \ 5\observerBandwidth^4 \ \observerBandwidth^5]^\top$ is a vector of observer gains.

% To match the rank of structure \eqref{eq:oscillatoryESO}, extended state utilized in the conventional ESO \eqref{eq:standardESO} needs to be augmented by another two state variables, making $\moreExtendedStateDerivative \triangleq [\controlError \ \controlErrorDerivative \ \totalDisturbance \ \totalDisturbanceDerivative \ \totalDisturbanceSecondDerivative]^\top$  a new extended state.

The rank of observer \eqref{eq:oscillatoryESO} is equal to $\rank=5$, so to effectively compare it with an observer designed in the same manner as \eqref{eq:standardESO}, we would like to introduce the new extended state vector $\moreExtendedStateDerivative \triangleq [\controlError \ \controlErrorDerivative \ \totalDisturbance \ \totalDisturbanceDerivative \ \totalDisturbanceSecondDerivative]^\top$, augmented by the additional two variables comparing to $\extendedState$, satisfying $\textrm{rank}(\moreExtendedState)=\textrm{rank}(\oscillatoryExtendedState)=5$.
The dynamics of $\moreExtendedState$ is expressed as
\begin{align}
    \begin{cases}
        \moreExtendedStateDerivative = \observerStateMatrix{5}\moreExtendedState - \frac{1}{\inertiaMatrixEstimate}\observerInputMatrix{5}(\controlSignal-\modeledDynamicsEstimate) + \disturbanceInputMatrix{5}\totalDisturbanceThirdDerivative \\
        \moreExtendedStateOutput = \observerOutputMatrix{5}\moreExtendedState + \measurementNoise
    \end{cases},
    \label{eq:moreExtendedStateDynamics}
\end{align}
where $\moreExtendedStateOutput$ is the observer output. Comparing to $\extendedState$ consisting of the combined vector $\combinedControlError$ and the total disturbance $\totalDisturbance$, vector $\moreExtendedState$ takes also the first and second derivatives of total disturbance as the additional states, inducing $\totalDisturbanceThirdDerivative$ to be a perturbation of \eqref{eq:moreExtendedStateDynamics}. The Luenberger ESO designed according to \eqref{eq:moreExtendedStateDynamics} is described as
\begin{align}
    \begin{cases}
        \moreExtendedStateEstimateDerivative = \observerStateMatrix{5}\moreExtendedStateEstimate - \frac{1}{\inertiaMatrixEstimate}\observerInputMatrix{5}(\controlSignal-\modeledDynamicsEstimate)+ \observerGainVector{5}(\moreExtendedStateOutput-\extendedObserverOutput)\\
        \extendedObserverOutput = \observerOutputMatrix{5}\moreExtendedStateEstimate
    \end{cases},
    \label{eq:extendedESO}
\end{align}
where $\observerGainVector{5}\triangleq[5\observerBandwidth \ 10\observerBandwidth^2 \ 10\observerBandwidth^3 \ 5\observerBandwidth^4 \ \observerBandwidth^5]^\top$ is a vector of observer gains.
%

% We would also like to present the comparison of the Luenberger-like observers \eqref{eq:standardESO}, \eqref{eq:oscillatoryESO}, and \eqref{eq:extendedESO} with the observers designed upon the corresponding dynamical equations according to the AM architecture presented in \cite{wang2017} that should result in the smaller amplification of high-frequency noise $\measurementNoise$ (as it was presented in \cite{astolfi2015}).
% All of the aforementioned observer structures are designed according to the Luenberger architecture. Besides changing the extended state definition between observers, we can also change the architecture of the designed ESOs. A technique of designing observers presented in \cite{wang2017} should result in the smaller amplification of the high-frequency noise measurement, especially for the systems with a high relative degree, according to the results presented in \cite{astolfi2015}. To be able to utilize AM observer design method, let us write down the generalized form of dynamics \eqref{eq:extendedStateDynamics}, \eqref{eq:moreExtendedStateDynamics}, and \eqref{eq:oscilatoryExtendedStateDynamics} as

All of the aforementioned observer structures are designed according to the Luenberger architecture. Besides changing the extended state definition between observers, we can also change the architecture of the designed ESOs using, for example, an Astolfi/Marconi observer design technique presented in \cite{wang2017}. According to \cite{astolfi2015}, the use of this method should result in the smaller amplification of the high-frequency noise measurement, especially for the systems with a high relative degree. To be able to utilize AM observer design method, let us write down the generalized form of dynamics \eqref{eq:extendedStateDynamics}, \eqref{eq:moreExtendedStateDynamics}, and \eqref{eq:oscilatoryExtendedStateDynamics} as
\begin{align}
    \begin{cases}
        \generalizedStateVectorElementDerivative{i} = \nonlinearStateFunctionElement{i}(\reducedGeneralizedStateVector{i},\generalizedStateVectorElement{{i+1}},\controlSignal) + \nonlinearDisturbanceInputFunctionElement{i}(\generalizedDisturbance), \ 1\leq i< n \\
        \generalizedStateVectorElementDerivative{n} = \nonlinearStateFunctionElement{n}(\reducedGeneralizedStateVector{n})+ \nonlinearDisturbanceInputFunctionElement{n}(\generalizedDisturbance)
    \end{cases},
\end{align}
where $\reducedGeneralizedStateVector{i} = [\generalizedStateVectorElement{1}, ... , \generalizedStateVectorElement{i}]^\top$ for $1\leq i\leq n$ and $(\generalizedStateVector \triangleq [\generalizedStateVectorElement{1},..., \generalizedStateVectorElement{n}]^\top = \extendedState, \ \generalizedDisturbance = \totalDisturbanceDerivative)$ for the dynamics \eqref{eq:extendedStateDynamics}, $(\generalizedStateVector = \moreExtendedState, \ \generalizedDisturbance = \totalDisturbanceThirdDerivative)$ for the dynamics \eqref{eq:moreExtendedStateDynamics} , and  $(\generalizedStateVector = \oscillatoryExtendedState, \ \generalizedDisturbance = \residualTotalDisturbanceDerivative)$ for the dynamics \eqref{eq:oscilatoryExtendedStateDynamics}. The Astolfi/Marconi observer, according to \cite{astolfi2015} and \cite{wang2017}, can be designed as
\begin{align}
    \begin{cases}
        \astolfiMarconiSubStateDerivative{i} &= \begin{bmatrix} \nonlinearStateFunctionElement{i}(\reducedGeneralizedStateVectorEstimate{i},\observerInputMatrix{2}^\top \astolfiMarconiSubState{i},\controlSignal) + \observerBandwidth\astolfiMarconiGainElement{i,1}\astolfiMarconiError{i} \\
        \nonlinearStateFunctionElement{{i+1}}(\reducedGeneralizedStateVectorEstimate{{i+1}},\observerInputMatrix{2}^\top \astolfiMarconiSubState{{i+1}},\controlSignal) +  \observerBandwidth^2\astolfiMarconiGainElement{i,2}\astolfiMarconiError{i} \end{bmatrix}, \ i\in\{1,...,n-2\} \\
        \astolfiMarconiSubStateDerivative{n-1} &= \begin{bmatrix} \nonlinearStateFunctionElement{{n-1}}(\reducedGeneralizedStateVectorEstimate{{n-1}},\observerInputMatrix{2}^\top \astolfiMarconiSubState{{n-1}},\controlSignal) + \observerBandwidth\astolfiMarconiGainElement{n-1,1}\astolfiMarconiError{{n-1}} \\
        \nonlinearStateFunctionElement{{n}}(\reducedGeneralizedStateVectorEstimate{{n}},\controlSignal) +  \observerBandwidth^2\astolfiMarconiGainElement{n-1,2}\astolfiMarconiError{{n-1}} \end{bmatrix} \\
        \reducedGeneralizedStateVectorEstimate{n} = \astolfiMarconiTransformationMatrix\astolfiMarconiState
    \end{cases},
    \label{eq:astolfiMarconiObserver}
\end{align}
where $\astolfiMarconiState \triangleq [\astolfiMarconiSubState{1}^\top \ ... \ \astolfiMarconiSubState{{n-1}}^\top]^\top$ is the observer state, $\astolfiMarconiGain{i} \triangleq [\astolfiMarconiGainElement{i,1},\astolfiMarconiGainElement{i,2}]^\top$ contains the design parameters,  while $\astolfiMarconiTransformationMatrix = \textrm{blkdiag}(\identityMatrix^2, \ \underbrace{\observerInputMatrix{2}, \ ..., \ \observerInputMatrix{2}}_{(n-2) \ \textrm{times}})$ is a transformation matrix between $\astolfiMarconiState$ and $\reducedGeneralizedStateVector{n}$. The estimation errors are defined as follows: $\astolfiMarconiError{1} \triangleq \controlError - \observerOutputMatrix{2}\astolfiMarconiSubState{1}$ and $\astolfiMarconiError{i} \triangleq \observerInputMatrix{2}^\top\astolfiMarconiSubState{{i-1}}-\observerOutputMatrix{2}\astolfiMarconiSubState{i}$ for $i\in\{2,...,n-1\}$.

\begin{remark}
    Observers \eqref{eq:standardESO} and \eqref{eq:oscillatoryESO} and their AM equivalents work properly when the signal $\totalDisturbanceDerivative$ is bounded, while observer \eqref{eq:extendedESO} and its AM equivalent need bounded $\totalDisturbanceThirdDerivative$. According to the dynamics \eqref{eq:errorDynamics}, controller \eqref{eq:controlLaw}, and assumption that the desired trajectory is at least $\desiredConfiguration\in\differentiableFunctionSet{5}$, we may claim that $\totalDisturbanceDerivative$ and $\totalDisturbanceThirdDerivative$ are bounded, as long as $\combinedControlError$ is in some compact set.
\end{remark}

%%%%%%%%%%%%%%%%%%%%%%%%%%%%%%%%%%%%%%%%%%%%%%%%%%%%%%%%%%%%%%%%%%
\section{Simulation results}

\begin{figure}[hbt!]
\vspace{-0.3cm}
 \centering
 \includegraphics[width=0.9\textwidth]{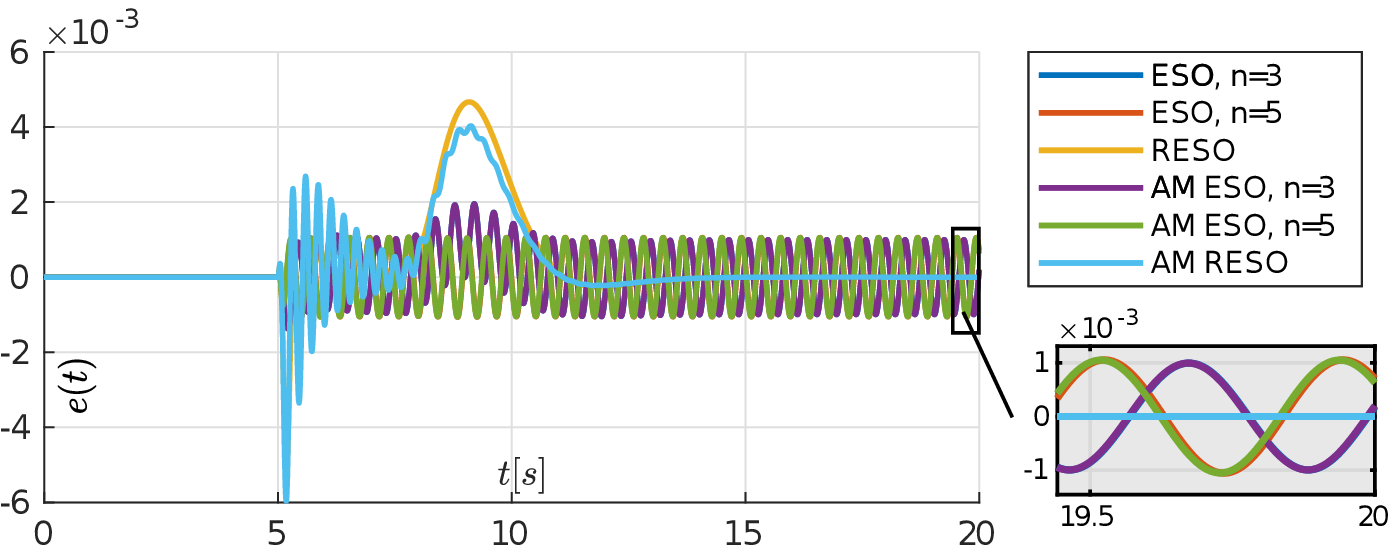} \\
 \includegraphics[width=0.9\textwidth]{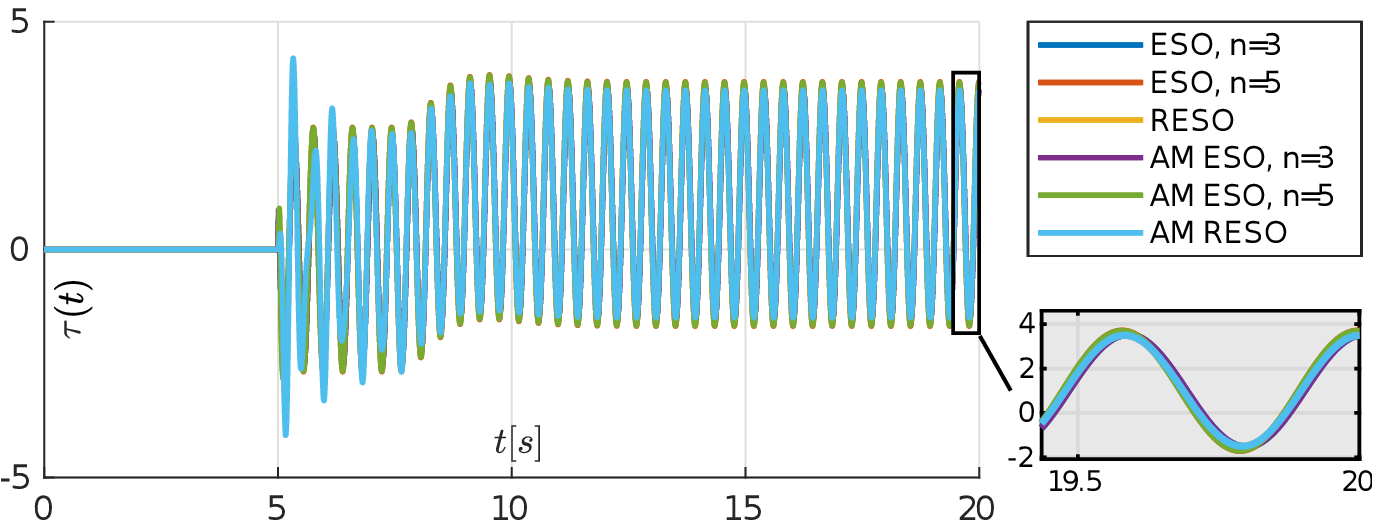}
 \vspace{-0.2cm}
 \caption{Control error and control signal for the case without measurement noise}
 \label{fig:1}
 \vspace{-0.0cm}
\end{figure}

\begin{figure}[hbt!]
 \centering
 \includegraphics[width=0.9\textwidth]{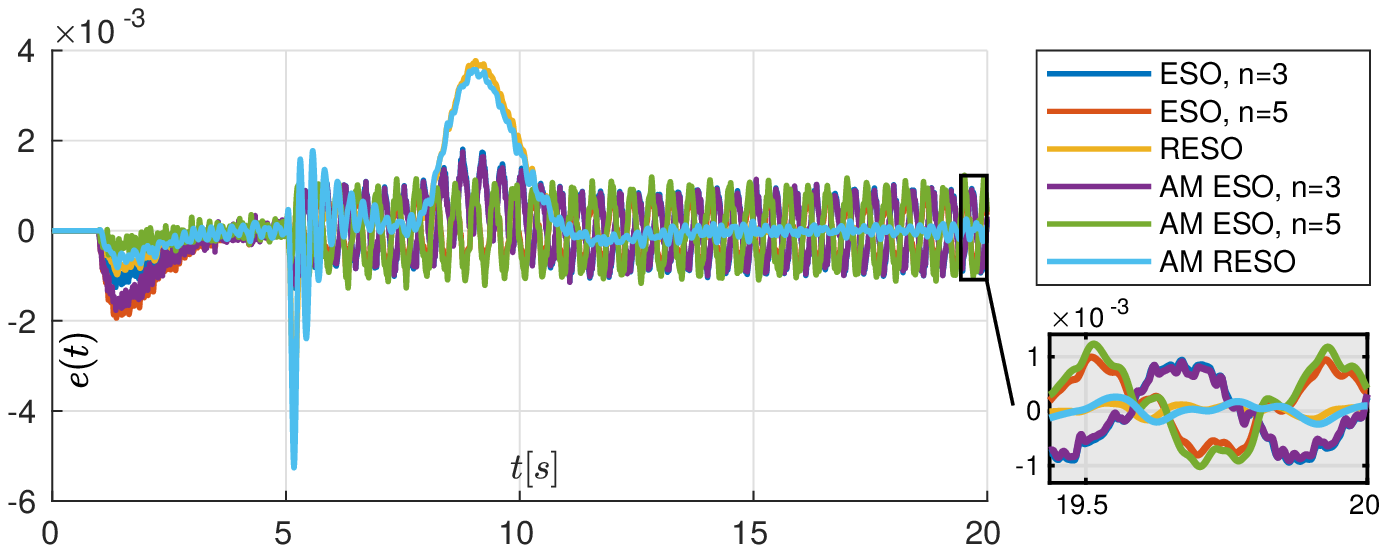} \\
 \includegraphics[width=0.9\textwidth]{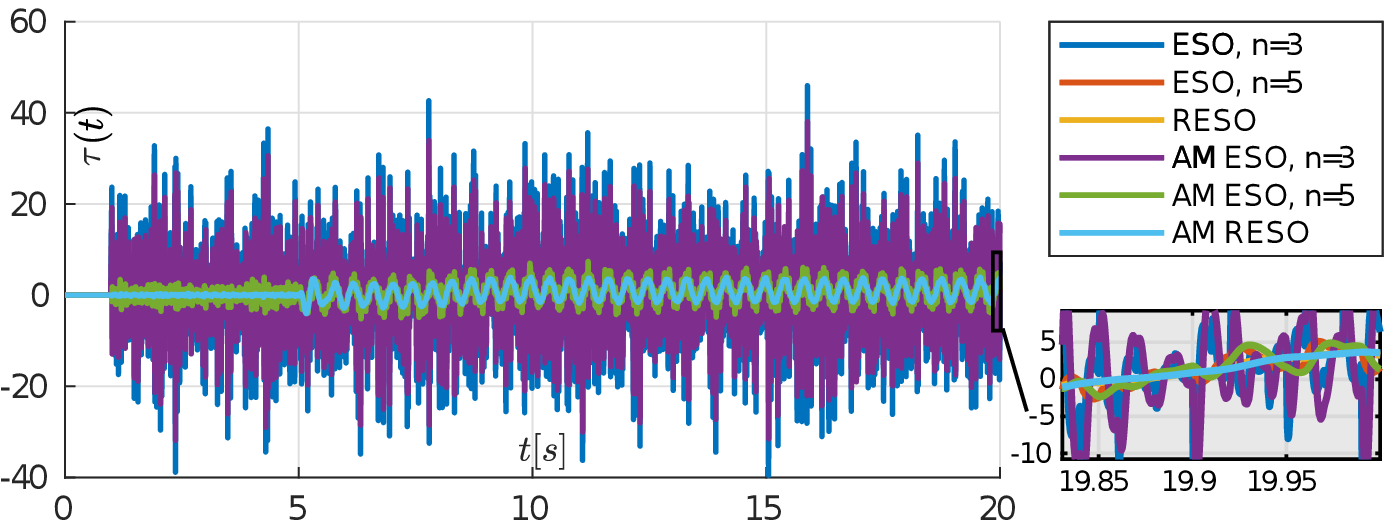}
 \vspace{-0.2cm}
 \caption{Control error and control signal for the case with measurement noise}
 \vspace{-0.3cm}
 \label{fig:2}
\end{figure}

The simulation study was conducted on the second order dynamical system described by transfer function $\transferFunction(\laplaceOperator) = 1/(\laplaceOperator+1)^2$. Two different scenarios have been considered and both concerned following the trajectory designed as the step function with a step time $t=7.5$ s and filtered with $\transferFunction_f(\laplaceOperator) = 1/(0.5\laplaceOperator+1)^5$ under the external disturbance $\externalDisturbance = 2.5\sin(15 t)$ applied after $t=5$ s. The first scenario presents the results without the measurement noise $\measurementNoise$, while the second one a case including white measurement noise with the variance $\sigma_w = 10^{-5}$.
According to the controller structure \eqref{eq:controlLaw}, we assumed that $\modeledDynamicsEstimate\equiv0$ and the parameters $\inertiaMatrixEstimate = 1$, $\proportionalGain=4$, $\differentialGain = 4$. In the presented results, we have used following abbreviations considering particular observers: ESO $n=3$ for \eqref{eq:standardESO}, ESO $n=5$ for \eqref{eq:extendedESO}, RESO for \eqref{eq:oscillatoryESO}, and we have added an AM prefix to refer to their equivalents designed with \eqref{eq:astolfiMarconiObserver}.
The values of $\astolfiMarconiGain{i}$ from \eqref{eq:astolfiMarconiObserver} were calculated with the tuning procedure presented in \cite{astolfi2015} resulting in ($\astolfiMarconiGain{1} = [0.8 \ 0.48]^\top, \ \astolfiMarconiGain{2}=[0.8 \ 0.16]^\top$) for AM ESO $n=3$ and ($\astolfiMarconiGain{1} = [0.6 \ 0.36]^\top, \ \astolfiMarconiGain{2} = [0.6 \ 0.135]^\top, \ \astolfiMarconiGain{3} = [0.6 \ 0.06]^\top, \ \astolfiMarconiGain{4} = [0.6 \ 0.025]^\top$) for AM ESO $n=5$ and RESO. A resonant angular frequency visible in \eqref{eq:harmonicOscillator} was set to $\disturbancePulsation=15$ rad/s, while the values $\observerBandwidth$ for all of the observers were chosen in a way to provide the same value of integral control cost criterion $\errorCost = \frac{1}{\simulationTime} \int^{\simulationTime}_0 \abs{\controlError(t)}dt$ for the simulation time $\simulationTime = 20$ s. Tables \ref{tab:1} and \ref{tab:2} contain the values of $\observerBandwidth$ and $\controlCost$ for all of the considered cases, together with the control cost $\controlCost = \frac{1}{\simulationTime} \int^{\simulationTime}_0 \abs{\controlSignal(t)}^2dt$ and the average quality of total disturbance estimation $\disturbanceErrorCost = \frac{1}{\simulationTime} \int^{\simulationTime}_0 \abs{\totalDisturbanceObservationError(t)}dt$. The values of control error $\controlError(t)$ and control signal $\controlSignal(t)$ are also presented in Figures \ref{fig:1} and \ref{fig:2}. To avoid a peaking phenomenon appearance in the figures, the controller is inactive during the observer transient stage and turned on after time $t= 1$ s (see \cite{huang2014}).

\begin{table}
\centering
\makebox[0pt][c]{\parbox{1.0\textwidth}{%
    \begin{minipage}[b]{0.49\hsize}\centering
        \vspace{-0.3cm}
        \caption{Results obtained for the first scenario, without measurement noise}
        \centering
        \vspace{-0.2cm}
        \begin{tabular}{labab}
            \rowcolor{White}
            \hline
            Observer type & \hspace{0.1cm} $\observerBandwidth$ \hspace{0.1cm}  & \hspace{0.1cm} $\errorCost$ \hspace{0.1cm}  & \hspace{0.1cm} $\controlCost$ \hspace{0.1cm} & \hspace{0.1cm} $\disturbanceErrorCost$ \hspace{0.1cm} \\
            \hline
            ESO $n=3$ & 490.03 & 0.01 & 59.77 & 2.41 \\
            ESO $n=5$ & 68.58 & 0.01 & 65.96 & 2.52 \\
            RESO & 27.32 & 0.01 & 59.31 & 1.05 \\
            AM ESO $n=3$ & 818.86 & 0.01 & 59.95 & 2.42 \\
            AM ESO $n=5$ & 340.27 & 0.01 & 65.31 & 2.54 \\
            AM RESO & 129.61 & 0.01 & 60.70 & 2.15  \\
            \hline
        \end{tabular}
        \vspace{-0.3cm}
        \label{tab:1}
    \end{minipage}
    \hfill
    \begin{minipage}[b]{0.49\hsize}\centering
        \vspace{-0.3cm}
        \caption{Results obtained for the second scenario, with measurement noise}
        \centering
        \vspace{-0.2cm}
        \begin{tabular}{labab}
            \rowcolor{White}
            \hline
            Observer type & \hspace{0.1cm} $\observerBandwidth$ \hspace{0.1cm}  & \hspace{0.1cm} $\errorCost$ \hspace{0.1cm}  & \hspace{0.1cm} $\controlCost$ \hspace{0.1cm} & \hspace{0.1cm} $\disturbanceErrorCost$ \hspace{0.1cm} \\
            \hline
            ESO $n=3$ & 586.76 & 0.01 & 623.71 & 69.25 \\
            ESO $n=5$ & 76.94 & 0.01 & 94.58 & 19.59 \\
            RESO & 31.52 & 0.01 & 60.06 & 3.31 \\
            AM ESO $n=3$ & 1057.46 & 0.01 & 844.89 & 92.91 \\
            AM ESO $n=5$ & 352.48 & 0.01 & 91.02 & 18.17 \\
            AM RESO & 140.34 & 0.01 & 60.41 & 3.10 \\
            \hline
        \end{tabular}
        \label{tab:2}
        \vspace{-0.3cm}
    \end{minipage}
}}
\end{table}

In the case without the measurement noise, selected values of $\observerBandwidth$ for Luenberger observers result in almost identical outcomes of the $\controlCost$ and $\disturbanceErrorCost$ values comparing to their AM equivalents (the only exception is the $\disturbanceErrorCost$ for AM RESO, see Table \ref{tab:1}, caused by a more dynamic transient stage that could be seen in the control error representation in Fig. \ref{fig:1}). A specific structure of RESO allows the control system to perform much better than ESO $n=5$ in the presence of sinusoidal disturbances, leading to the possibility of selecting smaller $\observerBandwidth$ values to obtain the same control quality $\errorCost$. As a consequence of lower $\observerBandwidth$, the level of measurement noise amplification in RESO is smaller than in the case of ESO $n=5$ (see $\controlCost$ values in Tables \ref{tab:1} and \ref{tab:2}).
 Control cost obtained for ESO $\rank=5$ and RESO is approximately the same for the Astolfi/Marconi and Luenberger observers (see a zoomed subplot of control signals in Fig. \ref{fig:2}).
 In the scenario with the measurement noise, the use of AM ESO $n=3$ observer results in a greater control cost comparing to its Luenberger equivalent, what may be an indication that this observer architecture works well for the systems with higher rank (at least in the case of the selected trajectory and comparison criterion).

%%%%%%%%%%%%%%%%%%%%%%%%%%%%%%%%%%%%%%%%%%%%%%%%%%%%%%%%%%%%%%%%%%
\section{Experimental validation}

\begin{wrapfigure}{R}{5.5cm}
	\centering
	\vspace{-0.6cm}
	\includegraphics[width=5.3cm]{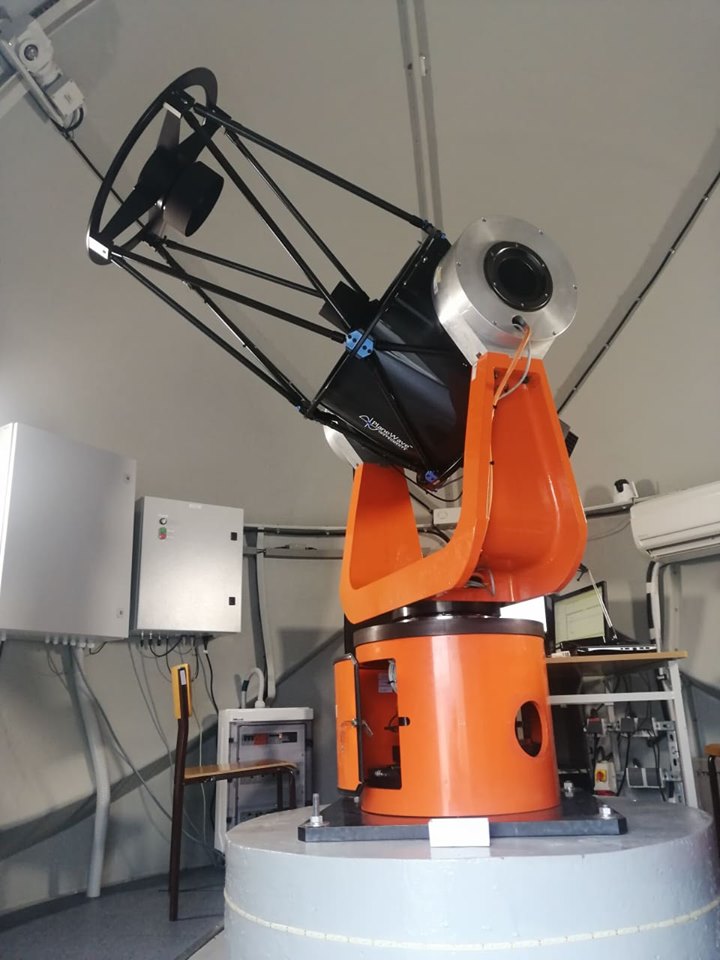}
	\vspace{-0.2cm}
	\caption{Astronomic telescope photo}
	\label{fig:telescope}
	\vspace{-0.6cm}
\end{wrapfigure}
\begin{figure}[hbt!]
 \centering
 \vspace{-0.3cm}
 \includegraphics[width=0.9\textwidth]{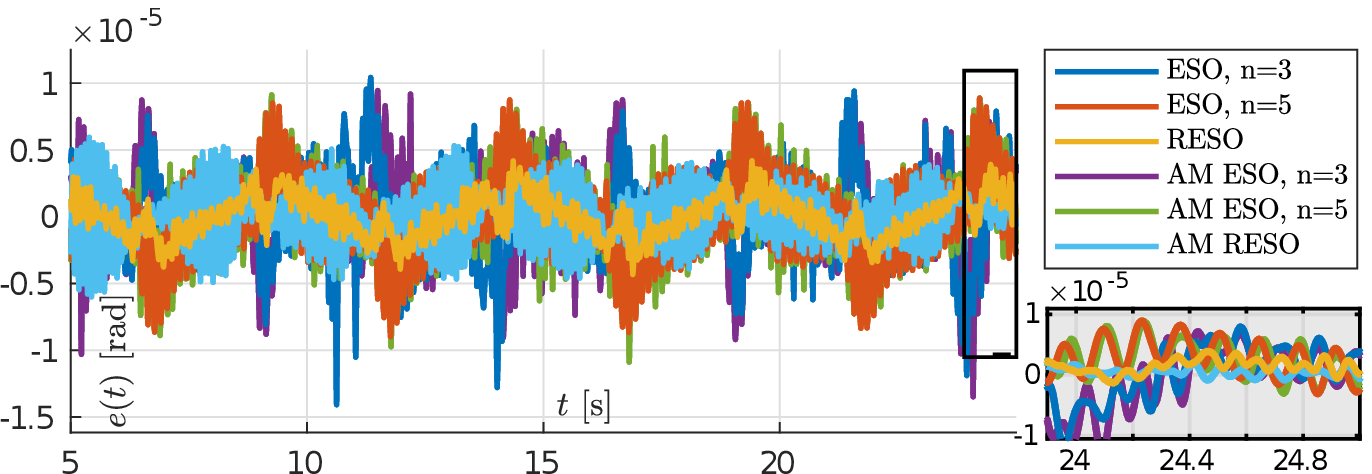} \\
 \includegraphics[width=0.9\textwidth]{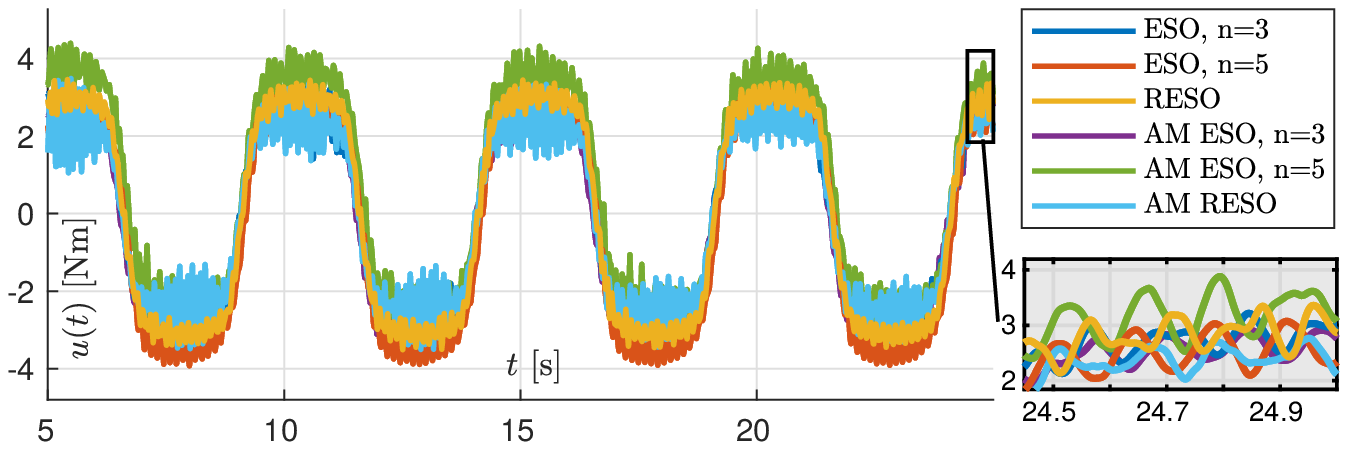} \\
 \includegraphics[width=0.9\textwidth]{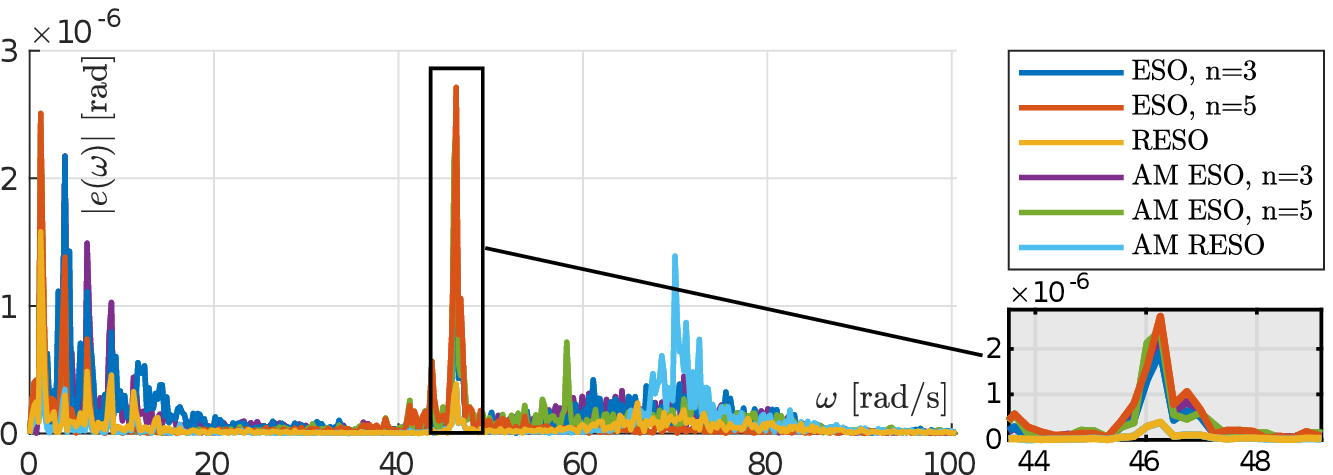}
 \vspace{-0.2cm}
 \caption{Tracking error, control signal and error spectrum obtained in the experiment}
 \vspace{-0.3cm}
 \label{fig:4}
\end{figure}

Experiments were conducted on the robotic mount of the astronomic telescope in a class 0.5-m that was presented in Fig. \ref{fig:telescope}. A kinematic structure of the mount can be described as a 2 degree-of-freedom manipulator with perpendicular revolute joints. Each joint is gearlessly driven by permanent-magnet synchronous motors whose angular position is measured using a high-precision 32-bit rotary encoder, see \cite{kozlowski2020}. Both motors are controlled by independent cascade controller with the position and current parts separated. The current loop of each axis contains a PI regulator and has much faster dynamics comparing to the loop associated with the mechanical part of the control system. In the experiments, we assume the influence of current control to be negligible and focus on the angular position control of the described system. To express the model of each telescope mount axis with the equation \eqref{eq:systemDynamics}, we assign $\configuration$ to be the angular position of a joint, $\controlSignal$ to the desired torque applied to the axis, while $\modeledDynamics$ represents friction forces estimated according to the LuGre model \cite{canudas1995} with parameter values identified in \cite{piasek2019}.
% Estimated parameters of the friction and identification procedure are as in \cite{piasek2019}.
In the experiments, we only considered the vertical (lower) axis of the mount, while the horizontal one was stabilized in a fixed upward position. The identified value of vertical axis inertia is equal to $\inertiaMatrixEstimate = 30 \unit{m \cdot s^2}$. Preliminary results for the most common ADR controller using a 3rd order ESO \eqref{eq:standardESO} resulted in the observation of the oscillatory disturbances with the angular frequency of $\omega_r \approx 46.9 \unit{rad}$ that was set up as resonant frequency of RESO and AM RESO observers. We performed the experiments for a slowly-varying sinusoidal reference trajectory $\desiredConfiguration = 2.89 \cdot 10^{-4} \sin (0.4 \pi t)$. Following such a slow trajectory is commonly required during astronomic observations (a maximal velocity of $\desiredConfiguration$ corresponds to the fivefold velocity of stars observed on a night sky) and demands an extremely precise control performance to be able to follow a chosen object. The tuning of all parameters was performed according to the comparison criterion chosen in the simulations. Values $\observerBandwidth$ for each observer were chosen to obtain comparable integral errors $\errorCost$ for all attempts. Due to the dominant
\begin{wraptable}{R}{6.5cm}
\centering
\vspace{-1.0cm}
    \caption{Results obtained in the experiments}
    \centering
    \vspace{0.1cm}
    \begin{tabular}{laba}
        \rowcolor{White}
        \hline
        Observer type & \hspace{0.1cm} $\observerBandwidth$ \hspace{0.1cm}  & \hspace{0.1cm} $\errorCost\ [10^{-6}]$ \hspace{0.1cm}  & \hspace{0.1cm} $\controlCost$ \hspace{0.1cm} \\
        \hline
        ESO $n=3$ & 140 & 2.60 & 6.39 \\
        ESO $n=5$ & 80 & 2.61 & 7.25 \\
        RESO & 80 & 1.20 & 7.16 \\
        AM ESO $n=3$ & 250 & 2.56 & 6.1 \\
        AM ESO $n=5$ & 450 & 2.62 & 7.03 \\
        AM RESO & 450 & 1.56 & 4.77 \\
        \hline
    \end{tabular} \\
    \label{tab:3}
    \vspace{-0.6cm}
\end{wraptable}
character of the oscillatory disturbance $\externalDisturbance$, it was not possible to obtain a similar control quality for RESO and ESO observers, so we have chosen the gains of RESO and RESO AM to be equal to ESO $n=5$ and its AM equivalent. Values of $\astolfiMarconiGain{i}$ were chosen as in the simulations. Obtained gains and values of $\errorCost$ and $\controlCost$ are presented in Table \ref{tab:3}. The results of the experiments are presented in Fig \ref{fig:4}. The initial conditions $\configuration(0)$ may strongly differ between particular experiments due to the very small range of angular positions generated by a trajectory generator (suitable for the astronomic observations), thus all plots are shown for $t \geq 5 \unit{s}$, when all of the transient stages have already passed.

The use of an extremely precise encoder resulted in the consistency between the outcome of experiments and simulations without introduced measurement noise for the Luenberger ESOs.  In the experiments, we obtained comparable control performance for 3rd and 5th order ESOs with a significantly lower gain value chosen for ESO $n=5$ followed by a slightly increased control cost $\controlCost$. Likewise, resonant observer designed according to \eqref{eq:oscillatoryESO} proved superior to the conventional variants resulting in much better compensation of sinusoidal perturbation visible in the control error frequency spectrum presented in Fig. \ref{fig:4}. Experiments performed with AM observers, although produced results similar to their Luenberger counterparts in the matter of chosen integral criteria, required different relative $\observerBandwidth$ values comparing to the simulation results and had in a slightly smaller values of $\controlCost$. While the parameters of AM ESO for $n=3$ were easily adjusted to provide desired performance, AM ESO $n=5$ required significantly higher $\observerBandwidth$ gain (despite the premises drew from simulations). Furthermore, AM RESO tended to cause unwanted vibrations around the resonant frequency of the mechanical structure (visible as a peak in the frequency spectrum of control error around $\omega\approx75 \unit{rad/s}$) leading to a destabilization of the control system for some gains $\observerBandwidth$.  Whether such behavior is inherent to discussed observer architecture or was caused by the unusual character of the plant remains yet uncertain. Still, for each of the AM observers, it was possible to obtain a satisfactory tracking quality comparable with its Luenberger counterpart by correctly choosing observer gain.

%%%%%%%%%%%%%%%%%%%%%%%%%%%%%%%%%%%%%%%%%%%%%%%%%%%%%%%%%%%%%%%%%%
\section{Conclusions}
Experimental and simulation results presented in this paper proved that it is possible to achieve the same control quality using different observer structures and architectures implemented as a part of the ADR controller designed for a mechanical system. Comparing to the 3rd and 5th order ESOs, the outcomes obtained with RESOs had smaller control cost values both for Luenberger and Astolfi/Marconi architectures when the system was perturbed with a significant sinusoidal disturbance. In the case of simulation with high measurement noise values, we have presented that the Luenberger architecture has better characteristics than the AM one for 3rd order ESOs, while for 5th order ESOs their outcome is comparable.

\bibliographystyle{plain}
\bibliography{bibliography}

\end{document}